\documentclass{PoS}

\usepackage{subfigure}

\title{PDFs and Neutrino-Nucleon Scattering from Hadronic Tensor}

\ShortTitle{Hadronic Tensor}

\author{\speaker{Jian Liang}\\
Department of Physics and Astronomy, University of Kentucky, Lexington, KY 40506, USA}

\author{Keh-Fei Liu\\
Department of Physics and Astronomy, University of Kentucky, Lexington, KY 40506, USA}

\abstract{
\begin{center}
\large{
\vspace*{0.4cm}
\includegraphics[scale=0.20]{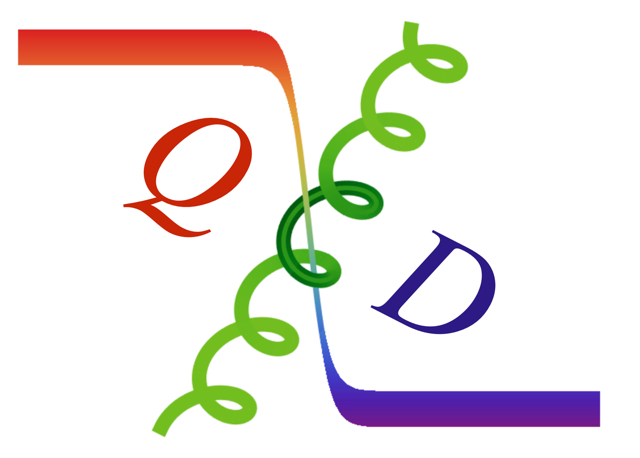}\\
\vspace*{0.4cm}
($\chi$QCD Collaboration) 
}
\end{center}
We review the Euclidean path-integral formulation of the nucleon hadronic tensor and classify the
gauge invariant and topologically distinct insertions in terms of connected and disconnected
insertions and also in terms of leading and higher-twist contributions in the DIS region. 
Converting the Euclidean hadronic tensor back to the Minkowski space requires solving an inverse problem
of the Laplace transform. We have investigated several inverse algorithms and studied the pros and
cons of each. 
We show a result with a relatively large momentum transfer ($Q^2 \sim 4\, {\rm GeV^2}$) 
to suppress the elastic scattering and reveal the contributions from the resonance and inelastic
region of the neutrino-nucleon scattering. For elastic scattering, the hadronic tensor is
the the product of the elastic form factors for the two corresponding currents. 
We checked numerically for the case of two charge vector currents ($V_4$) with the electric form factor 
calculated from the three-point function and found they agree within errors.
}

\FullConference{37th International Symposium on Lattice Field Theory - Lattice2019\\
16-22 June 2019\\
Wuhan, China}

\begin{document}

\section{Introduction}

Lepton-nucleon scattering cross-section is a product of the leptonic tensor and the hadronic
tensor. Being an inclusive reaction, the hadronic tensor includes all intermediate states
\begin{equation}   \label{w}
W_{\mu\nu}(q^2, \nu) = \frac{1}{2} \sum_n \int \prod_{i =1}^n \left[\frac{d^3 \vec{p}_i}{(2\pi)^3 2E_{i}}\right]  
\langle N|J_{\mu}(0)|n\rangle
\langle n|J_{\nu}(0) | N\rangle_{\rm spin\,\, ave.}(2\pi)^3 \delta^4 (p_n - p - q),
\end{equation}
where $p$ is the 4-momentum of the nucleon, $p_n$ is the 4-momentum of the nth intermediate state, and $q$ is the momentum transfer. 
Since $W_{\mu\nu}(q^2, \nu)$ measures the absorptive part of the  
Compton scattering, it is the imaginary part of the forward amplitude, and
can be expressed as the current-current correlation function in the nucleon, 
i.e.
\begin{equation}  \label{wcc}
W_{\mu\nu}(q^2, \nu) = \frac{1}{\pi} {\rm Im}\, T_{\mu\nu}(q^2, \nu)
= \int \frac{d^4x}{4\pi}  e^{ i q \cdot x}  \langle N| J_{\mu}(x)  J_{\nu}(0) | N\rangle_{\rm spin\,\, ave.}.
\end{equation} 
The hadronic tensor can be further decomposed, according to its Lorentz
structure, into structure functions, e.g.,
\begin{equation}
{W_{\mu\nu}}=\left(-g_{\mu\nu}+\frac{q_{\mu}q_{\nu}}{q^{2}}\right){F_{1}(x,Q^{2})}+\frac{\hat{p}_{\mu}\hat{p}_{\nu}}{p\cdot q}{F_{2}(x,Q^{2})}
\end{equation}
for the unpolarized case where $\hat{p}_{\mu}=p_{\mu}-\frac{p\cdot q}{q^{2}}q_{\mu}$.

\begin{figure}[ht]  
% \vspace*{1cm}
\centering
% \hspace{2cm}
\includegraphics[width=0.6\hsize]{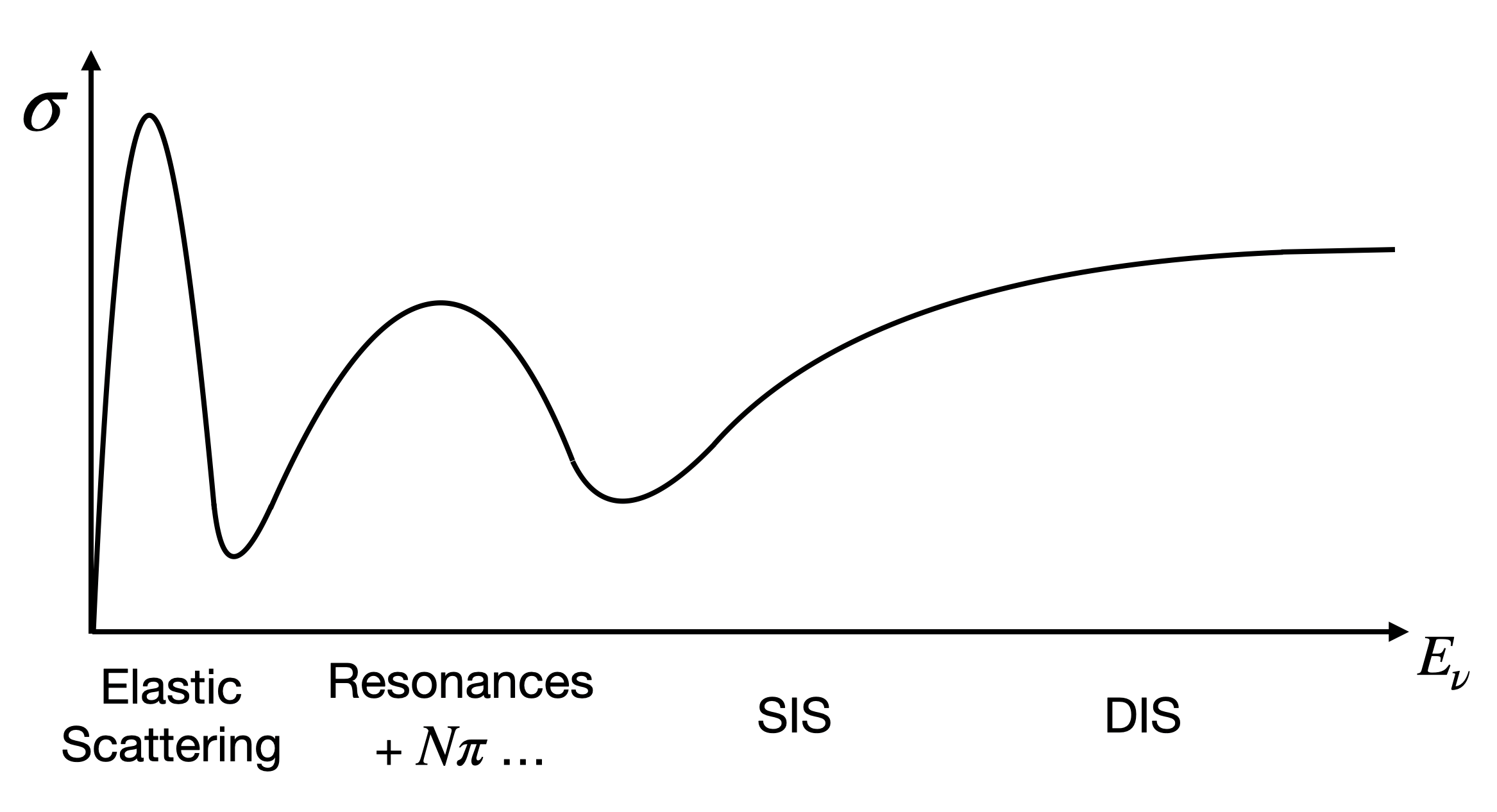}
\caption{Illustrated spectral density of the $\nu$-N scattering  to
show the elastic, the resonance, the SIS, and the DIS regions at different energy transfer $\nu$.  }
\label{nu-N-spectrum}
\end{figure}

The hadronic tensor is a function of $Q^2$ ($=-q^2$) and energy transfer $\nu$. 
The expected spectral density of the neutrino-nucleon scattering cross-section or structure
functions is illustrated in Fig.~\ref{nu-N-spectrum},
which shows that there are several kinematic regions in the spectral density in the energy 
transfer \mbox{$\nu$ -- the} elastic scattering, 
the inelastic reactions ($\pi N, \pi\pi N, \eta N$ etc.) and 
resonances ($\Delta$, Roper, $S_{11}$, etc.), 
the shallow inelastic scattering (SIS), and the deep inelastic scattering (DIS) regions. 
To determine how large a $\nu$ is needed for DIS, we look at $W$, the total invariant mass of the hadronic state 
\begin{equation}
W^2 = (q+p)^2 = m_N^2 - Q^2 + 2 m_N \nu,
\end{equation}
where $m_N$ is the nucleon mass.
The global fittings of the high energy lepton-nucleon and Drell-Yan experiments to extract 
the parton distribution functions (PDFs) usually make a cut with $W^2 > 10\, {\rm GeV^2}$. 
To be qualified in the DIS region, the energy transfer  $\nu$ needs to be 
\begin{equation}  \label{nu}
\nu\, > 4.86\, {\rm GeV} + 0.533\,(\rm GeV^{-1})\, Q^2.
\end{equation}
If we take  $Q^2 = 4\, {\rm GeV^2}$, this implies $\nu > 7$ GeV.
In the DIS region where the leading twists dominate, 
the cross-section or the structure functions can be used to extract 
the PDFs through the QCD factorization theorem
$F_{i}=\sum_{a}c_{i}^{a}\otimes f_{a}$, where the convolution kernel
$c_{i}^{a}$ is perturbatively calculable.

On the other hand, for scatterings at lower energies, 
the nucleon hadronic tenor is needed, 
together with many-body calculation of the nucleus, to delineate 
the experiments of neutrino-nucleus scattering, e.g.\ LBNF/DUNE~\cite{Acciarri:2015uup}
at Fermilab, which aims to study the neutrino properties. The beam
energy of DUNE is in the range $\sim1$ to $\sim7$ GeV and at different
beam energies, different contributions (quasi-elastic (QE), resonance
(RES), SIS and DIS) will dominate the total cross
section~\cite{Formaggio:2013kya}. The hadronic tensor is useful
in the whole energy region. For example in the QE region, the hadronic
tensor is actually the square of the elastic form factors of nucleon and the
cross-section of neutrino-nucleus scattering can be calculated by
combining the nucleon form factors and nuclear models about the nucleon
distribution inside a nucleus. In the RES, SIS and DIS region, inelastic
neutrino-nucleon scatterings emerge and one will need to have inclusive
hadronic tensor to cover the contributions of all final-state particles.
In this sense, calculating the hadronic tensor is so far the only
way we know that Lattice QCD can serve the neutrino experiments.

\section{Euclidean Path-integral Formulation of the Hadronic Tensor}  \label{HTparton}

The Euclidean hadronic tensor was formulated in the path-integral formalism 
to identify the origin of the Gottfried sum rule violation~\cite{Liu:1993cv}. 
It is a current-current correlator in the nucleon and can
be obtained by the following four-point-to-two-point correlator ratio
\begin{eqnarray}  \label{wmunu_tilde}
\widetilde{W}_{\mu\nu}(\vec{q},\vec{p},\tau) &=&
\frac{E_p }{m_N} \frac{{\rm Tr} (\Gamma_e G_{pWp}(t_0,t_1,t_2,t_f))}{{\rm Tr} (\Gamma_e G_{pp}(t_0,t_f))}
\begin{array}{|l} \\  \\  t_f -t_2 \gg 1/\Delta E_p, \, t_1 - t_0 \gg 1/\Delta E_p \end{array} \nonumber \\
&=& \left\langle N(\vec{p})| \int d^3\vec{x}  \frac{1}{4\pi} e^{-i\vec{q}\cdot \vec{x}}
J_{\mu}(\vec{x},\tau) J_{\nu}(0,0)|N(\vec{p})\right\rangle,
\end{eqnarray}
where $\tau = t_2 - t_1$ is the Euclidean time separation between the current $J_{\nu}(t_2)$ and 
$J_{\mu}(t_1)$. 
The source/sink of the nucleon interpolation fields at $t_0/t_f$ are much larger than the 
inverse of the energy difference between the nucleon and its first excited state ($\Delta E_p$), 
so that the nucleon excited states are filtered out. 
Formally, the inverse Laplace transform converts $\widetilde{W}_{\mu\nu}(\vec{q},\vec{p},\tau)$ to the Minkowski hadronic tensor
\begin{equation}  \label{wmunu}  
W_{\mu\nu}(\vec{q},\vec{p},\nu) = \frac{1}{2m_Ni} \int_{c-i \infty}^{c+i \infty} d\tau\,
e^{\nu\tau} \widetilde{W}_{\mu\nu}(\vec{q},\vec{p}, \tau),
\end{equation} 
with $c > 0$. This is basically doing the anti-Wick rotation to go back to the Minkowski space. 
In practice with the lattice calculation, 
it is not possible to perform the inverse Laplace transform in Eq.~(\ref{wmunu}), 
as there is no data at imaginary $\tau$. 
Instead, one can turn this into an inverse problem and find a solution from the Laplace transform~\cite{Liu:2016djw}
\begin{equation}  \label{Laplace}
\widetilde{W}_{\mu\nu} (\vec{q},\vec{p},\tau) = \int d \nu\, e^{-\nu\tau}
W_{\mu\nu}(\vec{q},\vec{p},\nu).
\end{equation}
This has been studied~\cite{Liang:2017mye,Liang:2019frk} with the inverse algorithms such as 
the Backus-Gilbert, Maximum Entropy, and Bayesian Reconstruction methods. 
\begin{figure}  
\centering
\subfigure[]
{{\includegraphics[width=0.30\hsize]{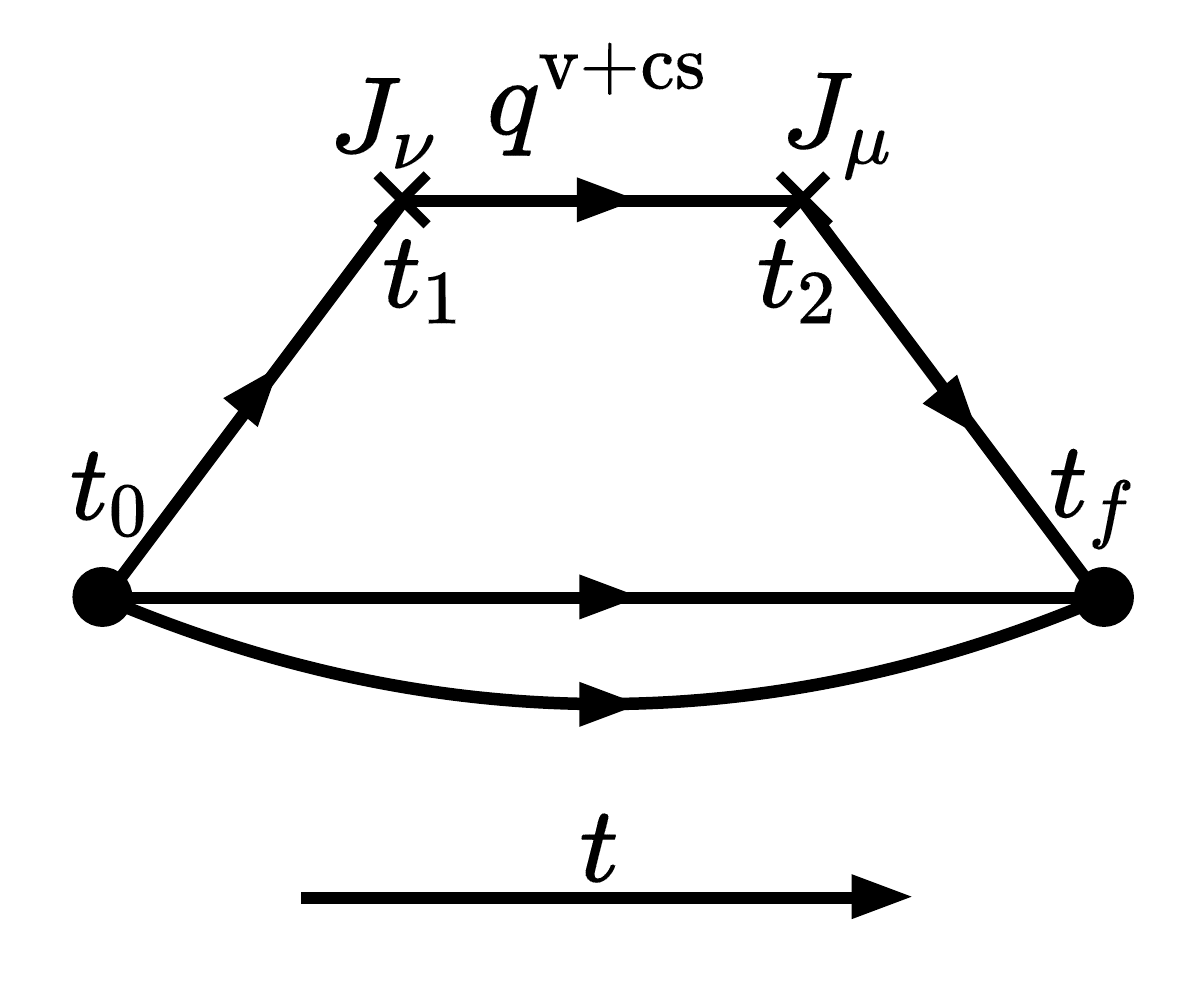}}
\label{v+CS}}
\subfigure[]
{{\includegraphics[width=0.30\hsize]{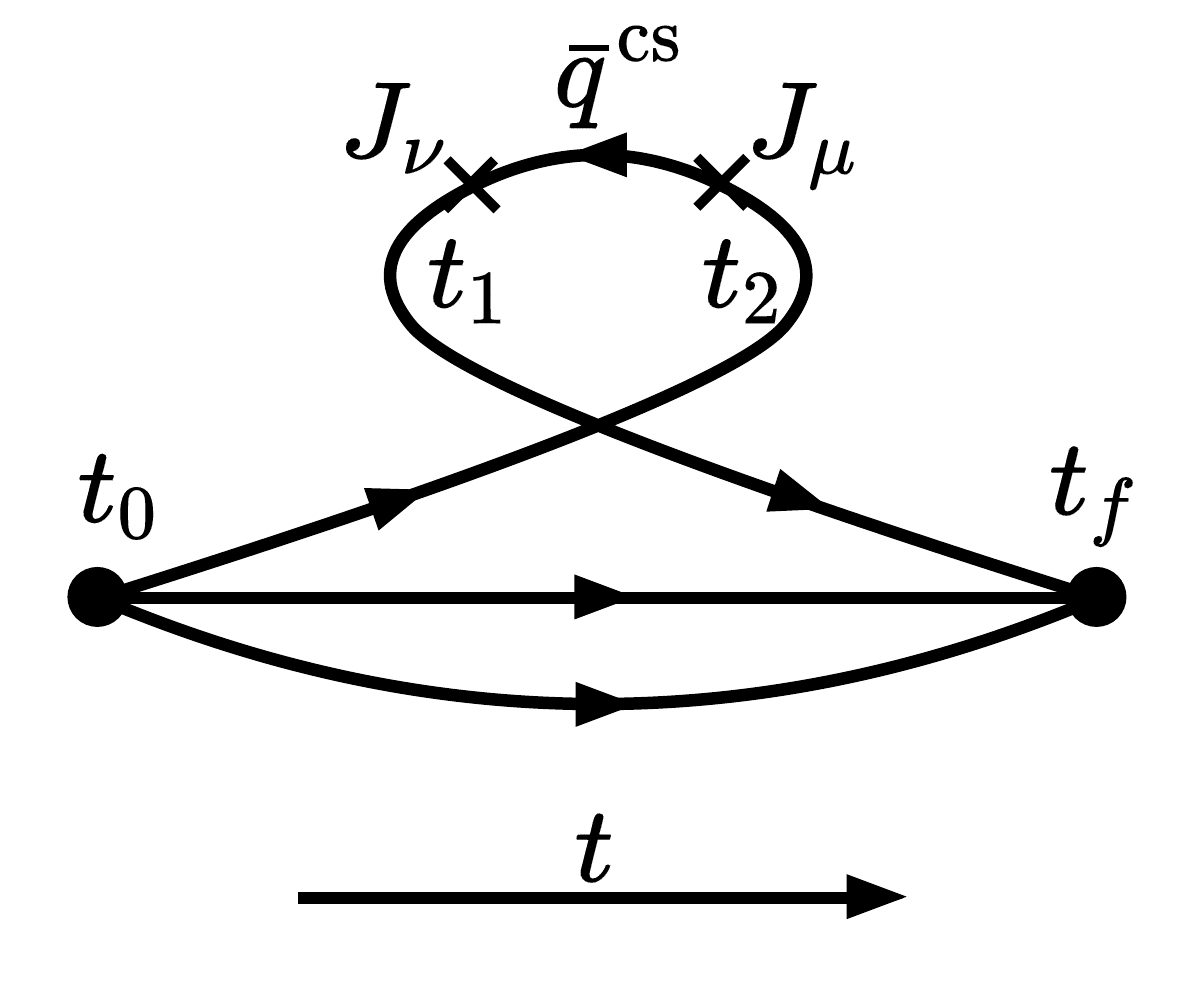}}
\label{CS}}
%\vspace*{-0.6cm}
\subfigure[]
{{\includegraphics[width=0.30\hsize]{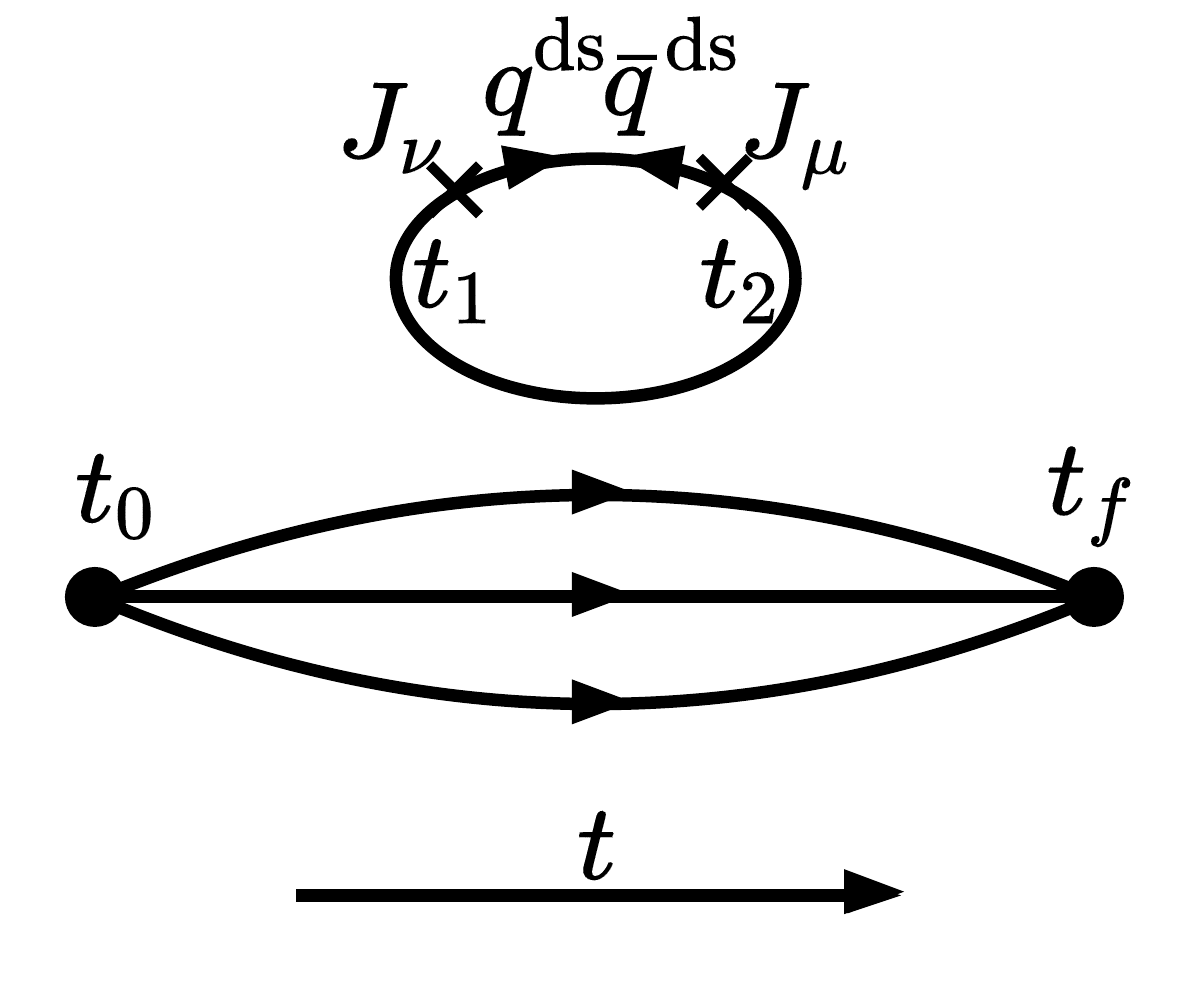}}
\label{DS}}
\caption{Three gauge invariant and topologically distinct insertions in the Euclidean path-integral
formulation of the nucleon hadronic tensor, where the currents couple to the same quark
propagator. In the DIS region, the parton degrees of freedom are
(a) the valence and connected sea (CS) partons $q^{v+cs}$, 
(b) the CS anti-partons $\bar{q}^{cs}$,
and (c) the disconnected sea (DS) partons $q^{ds}$ and anti-partons $\bar{q}^{ds}$ with $q = u, d, s,$ and $c$.  
Only $u$ and $d$ are present in (a) and (b) for the nucleon hadronic tensor. 
\label{leading-twist}}
%\newline
\vspace*{1cm}
\subfigure[]
{{\includegraphics[width=0.30\hsize]{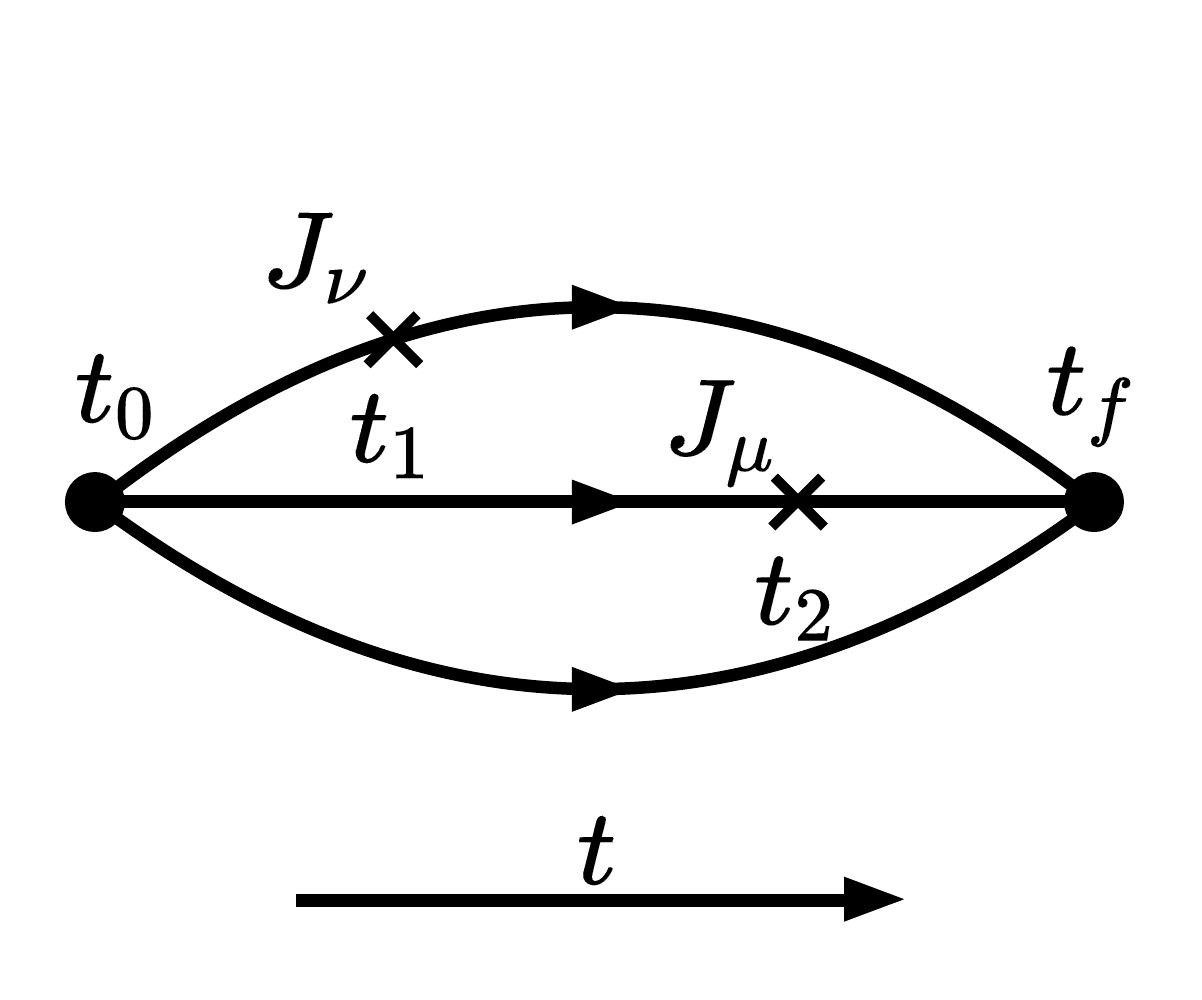}}
  \label{cat_ear_1}}
\subfigure[]
{{\includegraphics[width=0.30\hsize]{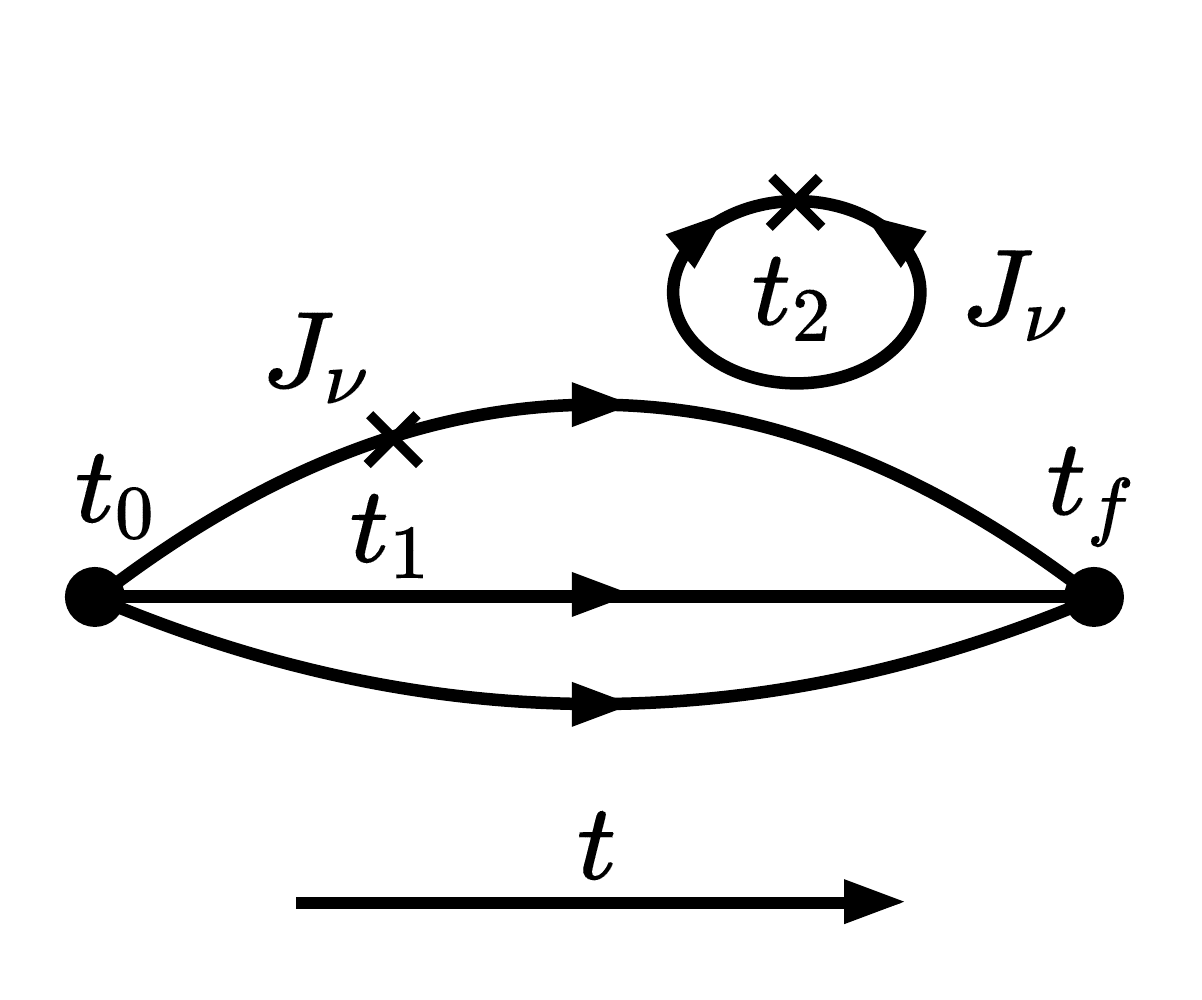}}
  \label{cat_ear_2}}
%\vspace*{-0.6cm}
\subfigure[]
{{\includegraphics[width=0.30\hsize]{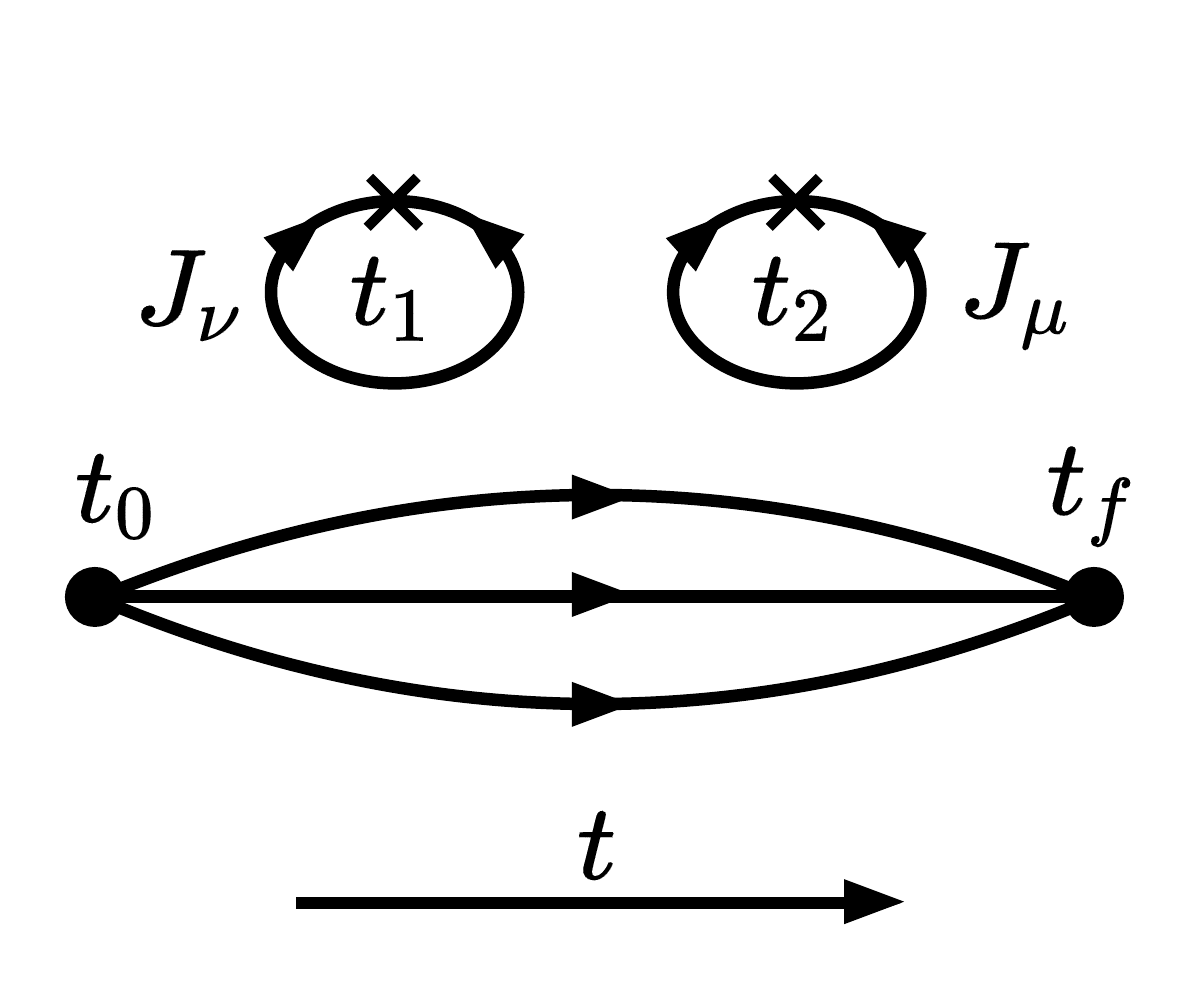}}
\label{cat_ear_3}}
\caption{Three other gauge invariant and topologically distinct insertions where the currents are
inserted on different quark propagators. In the DIS region, they contribute to higher twists.
\label{higher-twist}}
\end{figure}
It is shown~\cite{Liu:1993cv,Liu:1998um,Liu:1999ak} that, when the time ordering 
$t_f > t_2 > t_1 > t_0$ is fixed, the four-point function ${\rm Tr} (\Gamma_e G_{pWp}(t_0,t_1,t_2,t_f))$
can be grouped in terms of 6 topologically distinct and gauge invariant path-integral insertions 
as illustrated in Figs.~\ref{leading-twist} and \ref{higher-twist},
according to different Wick contractions among the Grassmann numbers in the two currents and the source/sink interpolation fields.  
They can be denoted as connected insertions (CI) (Fig.~\ref{v+CS} 
Fig.~\ref{CS}, and Fig.~\ref{cat_ear_1}) 
where the quark lines are all connected and disconnected insertions (DI) 
(Fig.~\ref{DS}, Fig.~\ref{cat_ear_2}, and Fig.~\ref{cat_ear_3}) 
where there is vacuum polarization associated with the current(s) 
in disconnected quark loop(s). 
Note, these diagrams depict the quarks propagating in the non-perturbative gauge background 
which includes the fluctuating gauge fields and virtual quark loops from the fermion determinant. 
Only the quark lines associated with the interpolation fields and currents are drawn in these quark skeleton diagrams. 

At low $\vec{q}$ and $\nu$ appropriate for a $\rho - N$ intermediate state, 
all CIs in Figs.~\ref{v+CS}, \ref{CS}, and \ref{cat_ear_1} 
contribute to the $\rho-N$ scattering. 
It is worth pointing out that Fig.~\ref{CS} includes the exchange contribution to prevent the $u/d$ quark in the loop in Fig.~\ref{DS} 
from occupying the same Dirac eigenstate in the nucleon propagator, 
enforcing the Pauli principle. In fact, Figs.~\ref{DS} and 
Fig.~\ref{CS} are analogous to the direct and exchange diagrams in time-ordered Bethe-Goldstone diagrams in many-body theory.

\subsection{Parton Degrees of Freedom}

In the DIS region (e.g.\ $Q^2 \ge 4\,{\rm GeV^2}$ and $\nu > 7$ GeV in Eq.~(\ref{nu})) as illustrated in Fig.~\ref{nu-N-spectrum}, 
insertions in Fig.~\ref{leading-twist} contain leading and higher twists (hand-bag diagrams), 
while those in Fig.~\ref{higher-twist} contain only higher twists (cat's ears diagrams). 
As far as the leading-twist DIS structure functions $F_1, F_2$, and $F_3$ are concerned,
the three diagrams in Fig.~\ref{leading-twist} are additive with contributions classified 
as the valence and sea quarks $q^{v+cs}$ in Fig.~\ref{v+CS}, the connected sea (CS) antiquarks $\bar{q}^{cs}$ in Fig.~\ref{CS}, 
and disconnected sea (DS) quark $q^{ds}$ and antiquarks $\bar{q}^{ds}$ 
in Fig.~\ref{DS}~\cite{Liu:1993cv,Liu:1999ak,Liu:1998um,Liu:2016djw}. 
It was pointed out that the Gottfried sum rule violation comes entirely from
the CS difference $\bar{u}^{cs} - \bar{d}^{cs}$ in the $F_2$ structure functions 
at the isospin symmetric limit~\cite{Liu:1993cv}. 

Owing to the factorization theorem~\cite{Collins:1989gx} 
which separates out the long-distance and short-distance behaviors, 
the structure function $F_1$ of the hadronic tensor can be factorized as
\begin{equation} \label{factorization}
F_1 (x, Q^2) = \sum_{i } \int_x^1 \frac{dy}{y}  C_i \Big(\frac{x}{y}, \frac{Q^2}{\mu^2}, \frac{\mu_f^2}{\mu^2}, \alpha_s(\mu^2)\Big)\, f_i (y, \mu_f, \mu^2),
\end{equation}
where $i$ is summed over quark $q_i$, anti-quark $\bar{q}_i,$ and glue $g$. 
$C_i$ are the Wilson coefficients and $f_i$ are the PDFs. $\mu_f$ is the factorization scale, 
and $\mu$ is the renormalization scale. In practice, 
the global fitting programs adopt the parton degrees of freedom as $u, d, \bar{u}, \bar{d}, s, \bar{s}$ and $g$. 
We see that from the path-integral formalism, each of the
$u$ and $d$ has two sources, one from the CI (Fig.~\ref{v+CS}) and one
from the DI (Fig.~\ref{DS}), so are $\bar{u}$ and $\bar{d}$ from Fig.~\ref{CS}
and Fig.~\ref{DS}. On the other hand, $s$ and $\bar{s}$ only come from the DI (Fig.~\ref{DS}).
In other words,
\begin{eqnarray}  \label{dof}
u\! &=&\! u^{v+cs} + u^{ds}, \hspace{2cm} d\,= \,d^{v+cs} + d^{ds} \nonumber \\
\bar{u}\! &=&\! \bar{u}^{cs} + \bar{u}^{ds}, \hspace{2.4cm}  \bar{d}\, = \,\bar{d}^{cs} + \bar{d}^{ds}, \nonumber \\
s\! &=&\! s^{ds},   \hspace{3.5cm} \bar{s}\, = \,\bar{s}^{ds}.
\end{eqnarray}

This classification of the parton degrees of freedom is richer than those in terms of $q$ and
$\bar{q}$ in the global analysis,
in which there are two sources for the partons -- $q^{v+cs}$ and $q^{ds}$ -- and two sources for the antipartons -- $\bar{q}^{cs}$ and $\bar{q}^{ds}$. They have different small
$x$ behaviors. For the CI part, $q^{v+cs}, \bar{q}^{cs} \sim x^{-1/2} $ where 
$q = u,d$; whereas, for the DI part, 
$q^{ds}, \bar{q}^{ds} \sim x^{-1} $ 
where $q = u,d,s,c$~\cite{Liu:1999ak,Liu:1998um,Liu:2012ch,Liang:2019xdx}. 
It is discerning to follow these degrees of freedom in moments which further 
reveals the roles of CI and DI in nucleon matrix elements. 
They have been intensively studied in lattice calculations which are beginning to take into account all the systematic corrections. 

\section{Large Momentum Transfer}

Numerically, a substantial challenge of this approach is to convert the hadronic
tensor from Euclidean space back to Minkowski space, which involves
solving an inverse problem of a Laplace transform~\cite{Liu:2016djw,Liang:2017mye,Liang:2019frk}. In order to tackle
this problem, three methods (i.e., the Backus-Gilbert method~\cite{Backus:1968bb,Backus:1970bb}, the
maximum entropy method~\cite{rietsch:1976ma,Asakawa:2000tr} and the Bayesian reconstruction method~\cite{Burnier:2013nla})
have been implemented and tested~\cite{Liang:2017mye,Liang:2019frk}. It is believed that the Bayesian
reconstruction method with further improvement~\cite{Fischer:2017kbq,Kim:2018yhk} is the best
method so far to solve the problem. On the other hand, the inverse problem
is a common problem in PDF calculations, and can be tackled by
using model-inspired fitting functions.
In order to study parton physics, another challenge of this approach is
to access high momentum and energy transfers such that the calculation can
cover the DIS region.   

It is shown recently in a lattice calculation that small lattice spacing (e.g.\ $a \le 0.04$ fm) 
is needed to reach such high energy excitation as $\nu \sim 7$ GeV in order to be in the DIS region as described
in Eq.~(\ref{nu})~\cite{Liang:2019frk}. The need of fine lattices appear to be also a
common challenge for the lattice PDF community. In order to alleviate excited-state contaminations on the source/sink side, 
one can consider variational calculations with multiple interpolation fields~\cite{Chen:2005mg}. 

\begin{figure}[!h]
  \centering
  \includegraphics[width=.48\textwidth,page=1]{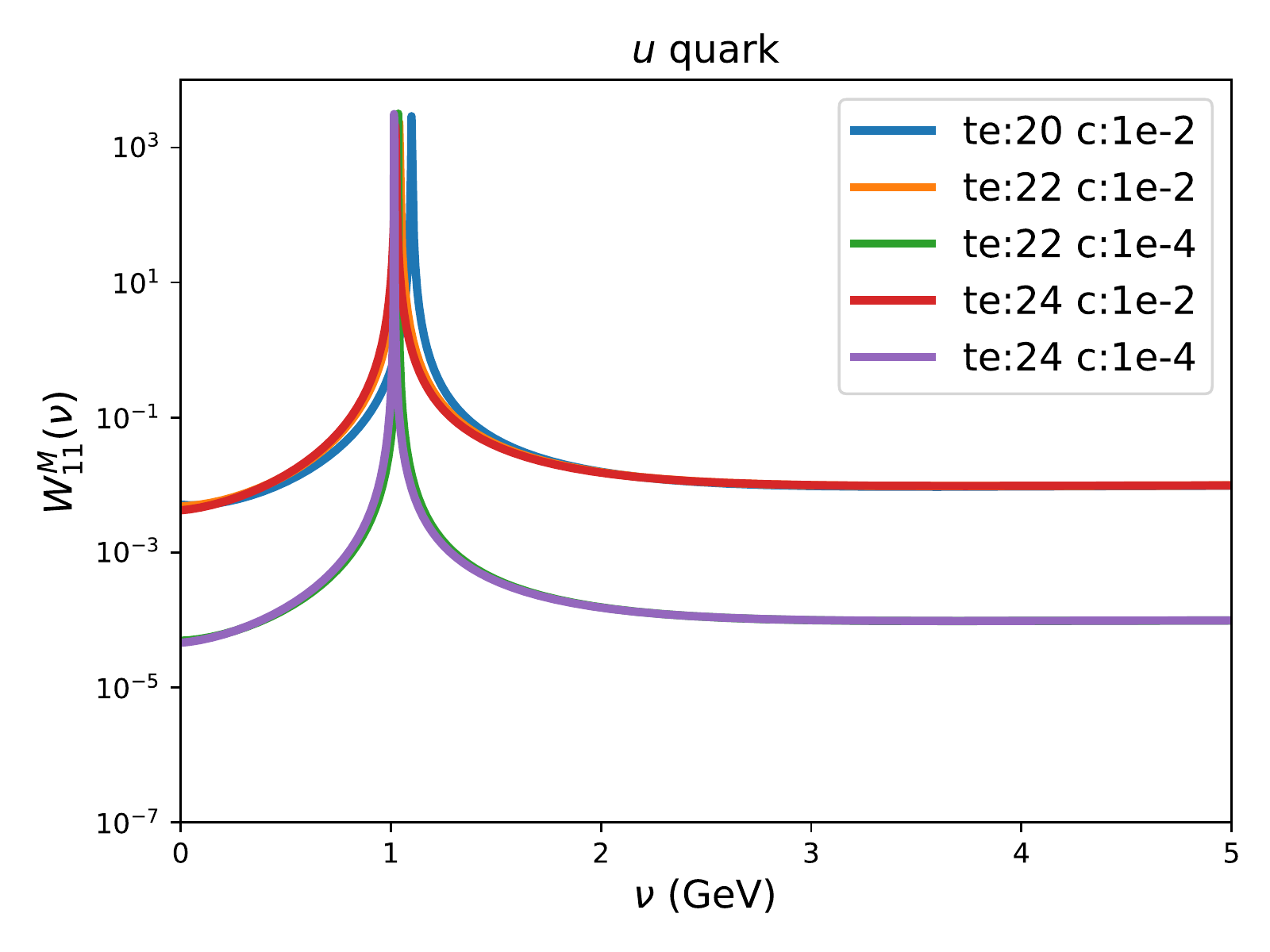}
  \includegraphics[width=.48\textwidth,page=1]{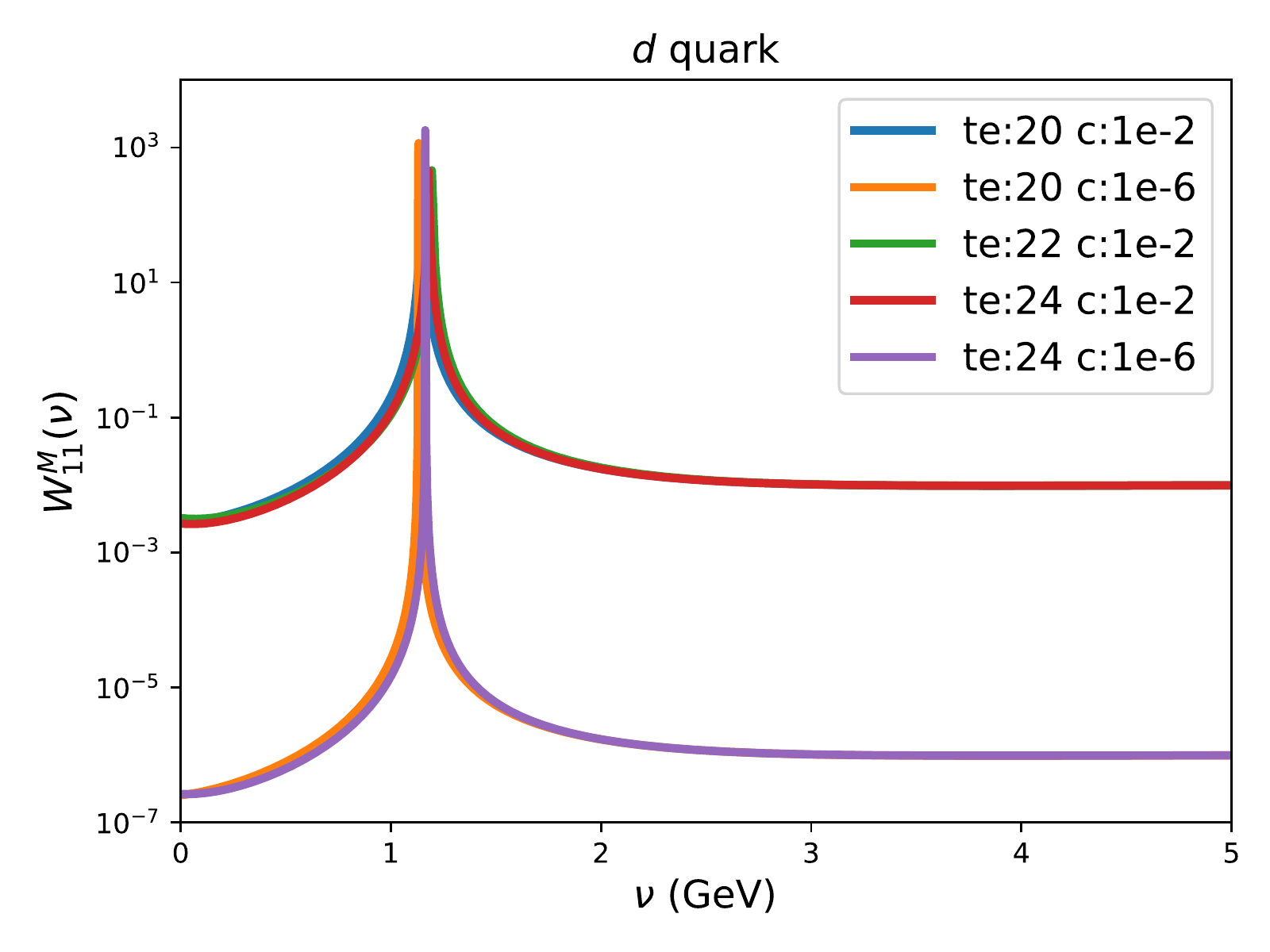}
  \caption{The Minkowski hadronic tensor ${W}^M_{11}$ as a function of energy transfer $\nu$ reconstructed 
  using the BR method for both $u$ and $d$ quarks.
  In the legend, ``$\rm te$'' denotes the end point of $t_2$ we use in BR 
  and ``$\rm c$'' is the value of the constant default model. After $\sim$ 2 GeV, the default models
  dominate the results.\label{M3}}
\end{figure}

We show the results in Fig.~\ref{M3} of a case with $Q^2 \sim 4\, {\rm GeV^2}$ so that the elastic scattering
contribution (at $\nu=0$) is suppressed. With Bayesian reconstruction, we see that there are substantial
contribution from the resonance and inelastic scattering at $\nu < 2$ GeV. After $\nu=$ 2 GeV, 
the default models
dominate the results, which is due to the fact that the lattice spacing of the ensemble used 
(clover lattice~\cite{Lin:2008pr} with $a_s=0.12$ fm, $\xi=3.5$) 
is not fine enough to
reveal higher intermediate states.

\section{Neutrino-Nucleon Scattering}

\begin{figure}[htbp]
  \centering
  \includegraphics[width=.5\textwidth,page=1]{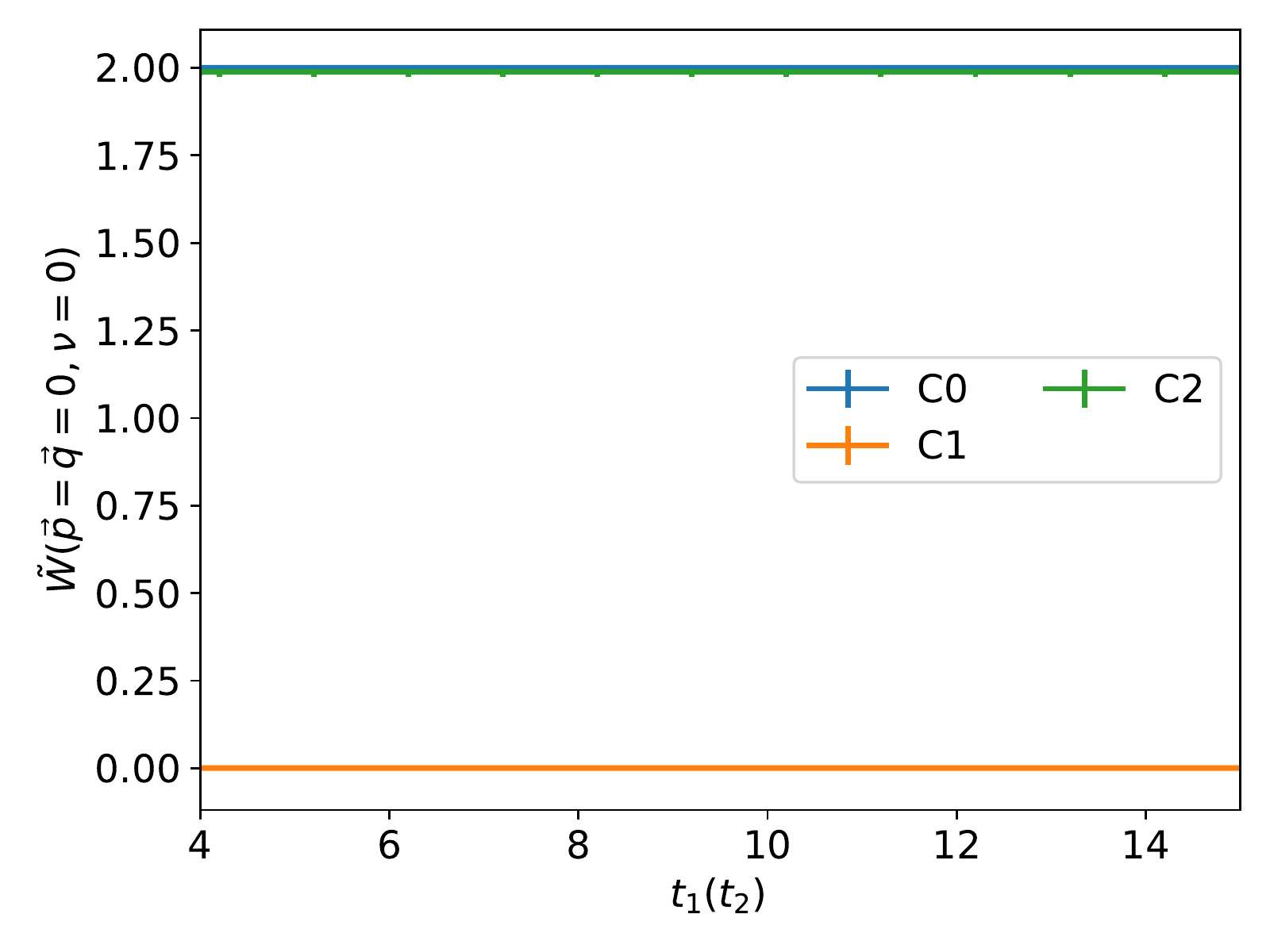}
  \caption{The contributions (CI only) of different insertions to the square of the total charges of 
  the $u$ quarks in the proton in the forward limit. Labels $\rm C0$, $\rm C1$, and $\rm C2$ in the figure denote
  the contributions from Fig.~\ref{v+CS}, Fig.~\ref{CS}, and Fig.~\ref{cat_ear_1}, respectively.
  The horizontal axis is the current position $t_1$ (for $\rm C0$ and $\rm C2$) or $t_2$ (for $\rm C1$).
  \label{C4-1}}
\end{figure}

For the hadronic tensor
calculation, one important point is that it works in all the energy ranges (from
elastic scattering to inelastic scattering and on to deep inelastic scattering). 
At low energy lepton-nucleon scattering, all the 6 diagrams in Figs.~\ref{leading-twist} and 
\ref{higher-twist} contribute and they are not separable.
For the elastic scattering case, the hadronic tensor $W_{\mu\nu}$ 
as a function of $Q^2$ is basically the product of the relevant nucleon form factors. 
For example, it is verified numerically (shown in Fig.~\ref{C4-1})
that it is the sum of Fig.~\ref{v+CS} and Fig.~\ref{cat_ear_1} 
that gives rise to the square of the total charges of the $u$ quarks in the proton 
in the forward limit when $J_{\mu}$ and $J_{\nu}$ are the charge currents, i.e.\
$W_{44} (\vec{p} = \vec{q} = 0, \nu = 0) (u\, {\rm quark}) = (2 e_u)^2$, 
while the other diagrams are zero due to charge conservation.

\begin{figure}[htbp]
  \centering
  \includegraphics[width=.32\textwidth,page=2]{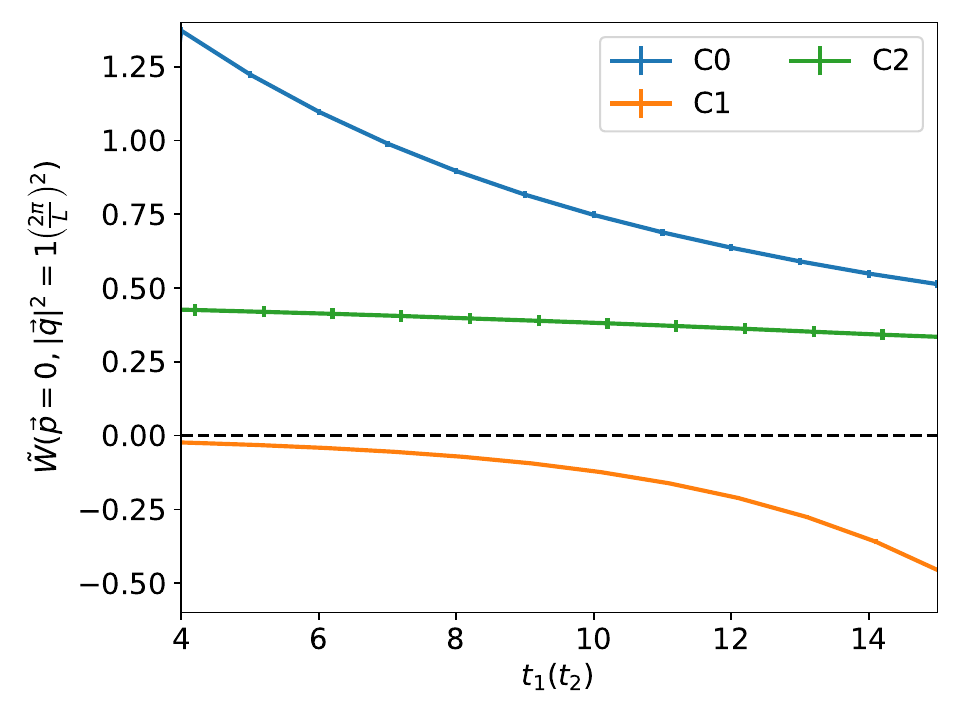}
  \includegraphics[width=.32\textwidth,page=3]{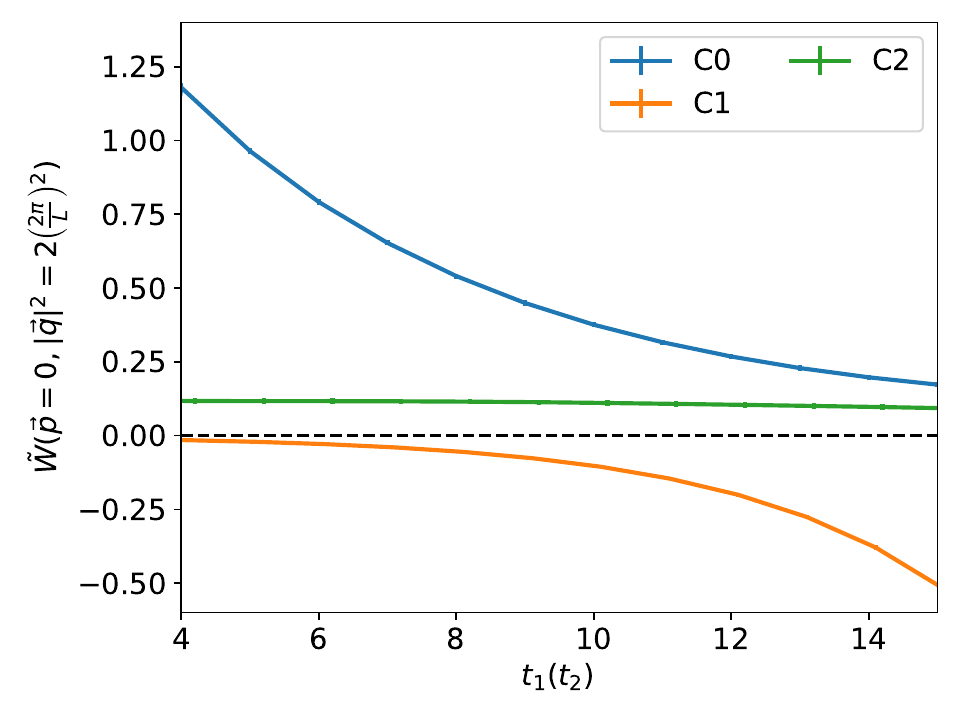}
  \includegraphics[width=.32\textwidth,page=4]{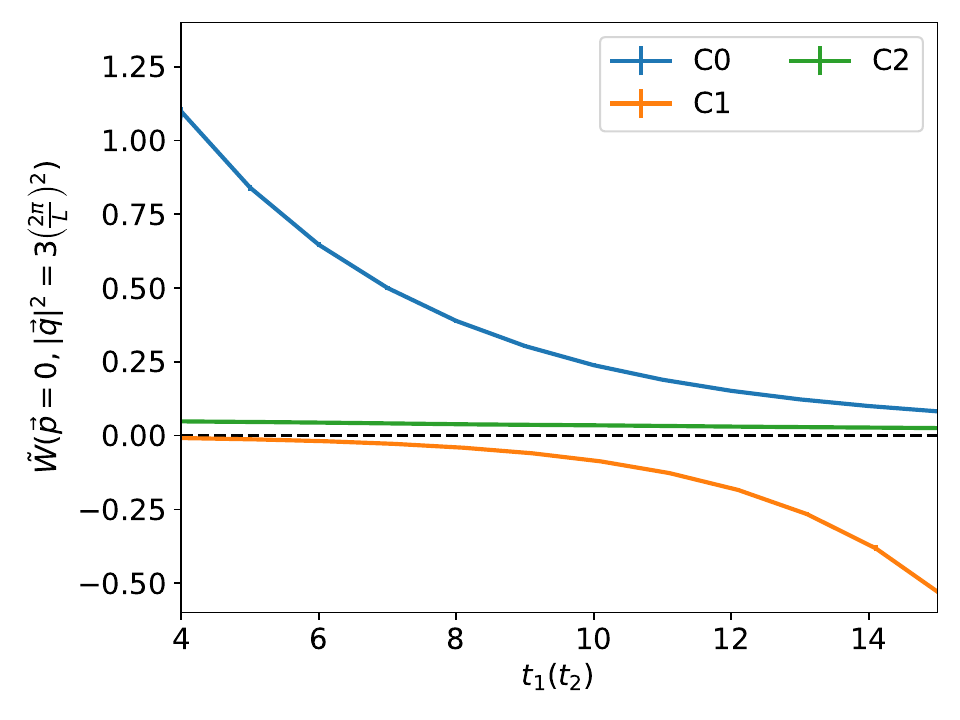}
  \caption{Similar to Fig.~\ref{C4-1} but for off-forward cases. The results of the first 3 momentum transfers
  are collected. A black dashed line of $\tilde{w}=0$ is added in each panel for legibility of the shift of the curves.
  \label{C4-2}}
\end{figure}

For the off-forward cases which are shown in Fig.~\ref{C4-2}, 
the contribution from Fig.~\ref{CS} (labelled as $\rm C1$ in the figure) is not zero,
as there is no charge symmetry in this case. A very interesting point 
is that the contribution from $\rm C2$ (Fig.~\ref{cat_ear_1}, the cat's ear diagram) becomes much smaller 
when the momentum transfer increases from $|\vec{q}|^2 = (2\pi/L)^2$ to $3 (2\pi/L)^2$ in the three panels in Fig.~\ref{C4-2},
while the contributions from $\rm C0$ and $\rm C1$ do not change much. This observation is consistent
with the fact that in DIS where the momentum transfer is large, the contribution from $\rm C2$ (Fig.~\ref{cat_ear_1})
is from the higher twist and is suppressed. Physically, this is because the two currents are acted on different quark lines
and extra gluon exchanges are needed for the off-forward matrix element of the hadronic tensor. 
Thus, the calculation of the hadronic tensor on the lattice
can be used to study the higher-twist effects explicitly.

\begin{figure}[htbp]
  \centering
  \includegraphics[width=.5\textwidth,page=5]{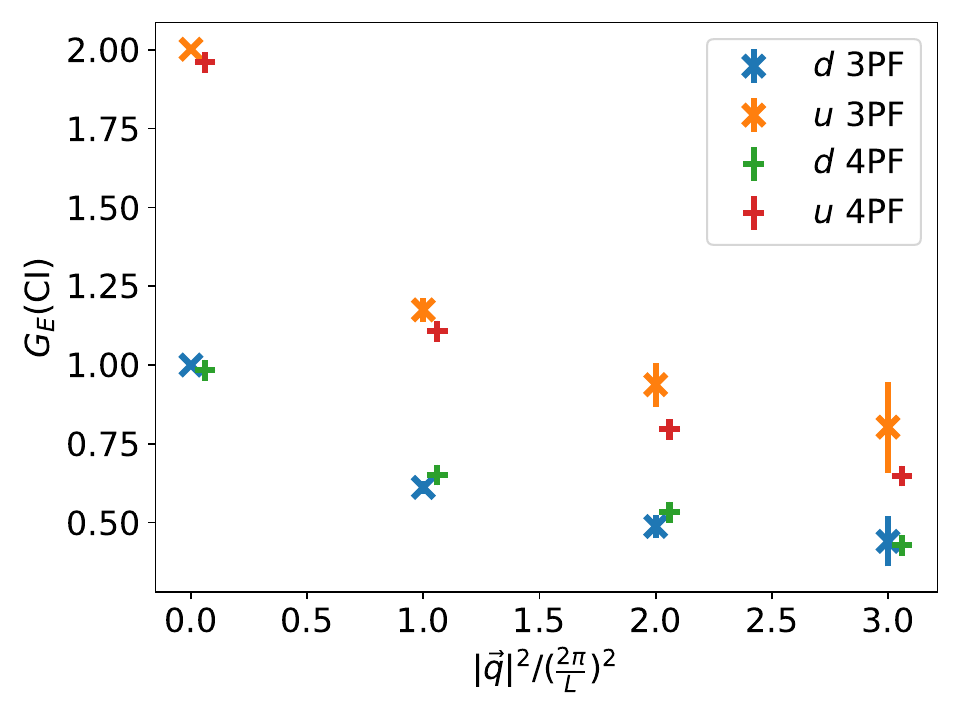}
  \caption{The comparison of the electric form factors (CI contributions only) calculated by using three-point functions and
  four-point functions for the first four momentum transfers (including zero) and for both $u$ and $d$ quarks.
  \label{check}}
\end{figure}

All the numerical check of this approach in terms of nucleon electric form factors
are done
on the $\rm RBC/UKQCD$ 32Ifine lattice~\cite{Blum:2014tka}.
The structure function of the elastic scattering from
the hadronic tensor, a four-point function, is the product
of the elastic nucleon form factors for the currents
involved. Fig.~\ref{check} shows that the electric form factors 
(connected insertions only) calculated by means of the
three-point functions for both $u$ and $d$ quarks are found to
be consistent within errors with those deduced from the
hadronic tensor for the elastic scattering.
This lays a solid foundation for further calculations of the neutrino scattering cross-sections.

\section{Summary and Outlook}

We formulate the hadronic tensor on the lattice and point out the
fact that this approach works for scattering processes at all
energy regions. We show a result with a relatively large momentum transfer 
which reveals the resonance and inelastic-scattering contributions in neutrino-nucleon scattering.
For elastic scattering,
we checked numerically for the case of two charge vector currents ($V_4$) with the electric form factor 
calculated from the three-point function and found they agree within errors. 

Currently, lattices with lattice spacing $\sim$0.04 fm are
suitable, for instance, to study the neutrino-nucleus
scattering for DUNE experiments, where the neutrino
energy is between $\sim$1 to $\sim$7 GeV. At the same
time, these lattices can also be used to explore the
pion PDFs. In the future, working on lattices with lattice
spacing of $\sim$0.03 fm or even smaller would be desirable for
studying the nucleon substructure in the DIS region.

\section{Acknowledgement}
This work is partially support by the U.S. DOE grant DE-SC0013065 and DOE Grant No.\ DE-AC05-06OR23177 which is within the framework of the TMD Topical Collaboration.
This research used resources of the Oak Ridge Leadership Computing Facility at the Oak Ridge National Laboratory, which is supported by the Office of Science of the U.S. Department of Energy under Contract No.\ DE-AC05-00OR22725. This work used Stampede time under the Extreme Science and Engineering Discovery Environment (XSEDE), which is supported by National Science Foundation Grant No. ACI-1053575.
We also thank the National Energy Research Scientific Computing Center (NERSC) for providing HPC resources that have contributed to the research results reported within this paper.

%\bibliographystyle{unsrt}
%\bibliography{trace_anomaly.bib}

\end{document}